\documentclass[manuscript]{aastex}

\usepackage{graphicx}

\begin{document}                                                                     

      
\title{Resolved Companions of Cepheids: Testing the Candidates with X-Ray
Observations\footnote{Based on observations obtained with {\it XMM-Newton},
an ESA science mission with instruments and contributions directly funded by
ESA Member States and the USA (NASA).}
}                                                                 


\author{Nancy Remage Evans,\altaffilmark{1}
Ignazio Pillitteri,\altaffilmark{1,2}
Scott Wolk,\altaffilmark{1}
Margarita Karovska,\altaffilmark{1}
Evan Tingle,\altaffilmark{1}  
Edward Guinan,\altaffilmark{3}
Scott Engle,\altaffilmark{3}
Howard E. Bond,\altaffilmark{4,5}
Gail H. Schaefer,\altaffilmark{6} 
and
Brian D. Mason\altaffilmark{7} }
   

\altaffiltext{1}
{Smithsonian Astrophysical Observatory,    
MS 4, 60 Garden St., Cambridge, MA 02138; nevans@cfa.harvard.edu}

\altaffiltext{2}
{INAF-Osservatorio di Palermo, Piazza del Parlamento 1, I-90134 Palermo, Italy}

\altaffiltext{3}
{Department of Astronomy and Astrophysics, Villanova University, 800 Lancaster Ave., 
Villanova, PA, 19085, USA}

\altaffiltext{4}
{Dept. of Astronomy \& Astrophysics, Pennsylvania State University,
University Park, PA 16802; heb11@psu.edu}

\altaffiltext{5}
{Space Telescope Science Institute, 3700 San Martin Drive, Baltimore, MD
21218}

\altaffiltext{6} {The CHARA Array of Georgia State University, Mount Wilson, 
California 91023, USA;  schaefer@chara-array.org}

\altaffiltext{7} {US Naval Observatory, 3450 Massachusetts Ave., NW, 
Washington, DC 20392-5420 }





\begin{abstract}

We have made {\it XMM-Newton\/} observations of 14 Galactic Cepheids that have
candidate resolved  ($\geq$5$\arcsec$) companion stars
based on our earlier {\it HST\/} WFC3
imaging survey. Main-sequence stars that are young enough to be physical
companions of Cepheids are expected to be strong X-ray producers in contrast to 
field stars. {\it XMM-Newton\/} exposures were set to detect 
essentially all companions hotter than spectral type M0 (corresponding to
0.5 $ M_\odot$.)
The large majority of our candidate companions were not detected in
X-rays, and hence are not confirmed as young companions. One  resolved
candidate  (S~Nor \#4)
 was unambiguously detected, but the Cepheid is a
member of a populous cluster. For this reason, 
 it is
likely that S~Nor \#4 is a cluster member rather than a gravitationally bound
companion. Two further Cepheids (S~Mus and R~Cru) have X-ray emission that
might be produced by either the Cepheid or the candidate resolved companion. 
A subsequent {\it Chandra} observation of S Mus shows that the X-rays 
are at the location of the Cepheid/spectroscopic binary.  
R Cru and also 
 V659 Cen (also X-ray bright) have possible companions closer than  5$\arcsec$ (the
limit for this study) which are the likely source of  X-rays. One final X-ray
detection (V473 Lyr) has no known optical companion, so the prime  suspect is
the Cepheid itself. It is a unique Cepheid with a variable amplitude. 


The 14 stars that we observed with {\it XMM\/} constitute 36\% of the 39
Cepheids found to have candidate companions in our {\it HST}/WFC3 optical
survey. No young probable binary companions were found with separations of $\geq\!5''$
or 4000 AU.



\end{abstract}




\section{Introduction}

Binary/multiple configurations influence every phase of
a star's life: formation,  the main sequence, and
the post-main sequence, and often drastically affect the outcomes.
For instance, Kraus et al.
(2012) find that binaries with intermediate separations (40 AU)
are less likely to have protoplanetary disks than both closer and wider
binaries, and hence are less likely to form planets.
For main-sequence stars, Sana et al.\ (2012) recently showed
that more than 70\% of O  stars ($\geq\!20\, M_\odot$) are destined to 
undergo binary interactions with mass exchange, and a third of those
undergo mergers. At later stages, many classes of X-ray sources 
have scenarios involving binaries
containing a compact object (supernovae progenitors, X-ray
binaries, novae, cataclysmic variables, and symbiotic stars). 
However, there  are many unanswered questions about 
how these configurations are achieved.  

Technical developments in recent years
in both radial velocities and imaging have contributed
greatly to the studies of binary/multiple stars.  For massive stars, the
velocity study of O stars in the 30 Dor (Tarantula 
Nebula) in the LMC (Sana et al.\ 2013) is of particular importance. It
confirmed the large fraction 
of binaries, and demonstrates the importance of interaction
inat the   their evolution (above; Sana et al.\ 2012).  A comparable velocity survey
was done on the B stars in the Tarantula Nebula (Dunstall et al.\ 2015), who
concluded that the multiplicity properties of the B stars are essentially the 
same as for the O stars.  An extensive high-angular-resolution study of O stars  
was made at the  Very Large Telescope (VLT) and
 the  Very Large Telescope Interferometer (VLTI) which 
covers separations from 0.5 to 10$^4$ mas (1 to 20,000 AU at a typical distance of 2 kpc;
Sana et al.\ 2014). Of particular importance was filling 
 the gap between
spectroscopic binaries and visual binaries 
which had been found by Mason et al.\ (1998). 
  A further survey was made by Aldoretto et al.\ (2015), of 
O and B stars using the {\it HST\/} Fine Guidance Sensor (FGS).  
A summary of previous studies of binarity/multiplicity
in massive stars can be found in Evans et al.\ (2013) and
Evans et al.\ (2016a).

Binary/multiple properties reflect the processes of star formation in 
their extent and mass ratios.  
The formation of wide binaries (1000 AU or greater) is a problem for 
star formation, since this is larger than the typical size of a star-forming core. 
Several mechanisms have been suggested to get around this problem. 
One  is the ``unfolding'' of a 
triple system.  In a triple which is dynamically unstable, one 
component (typically the smallest) may be sent to a wider orbit
(Reipurth \& Mikkola 2012).  
Another mechanism is the acquisition of a third star during cluster dissolution
(Kouwenhoven et al.\ 2010).   Parker \& Meyer (2014)
concluded that over time the reduction of triple systems through
disruption is balanced by acquisition of another star, keeping 
the binary fraction relatively constant.


The stars targeted in the present study are wide resolved companions.  Within 
such systems there is ample room for a close binary, and hence wide
companions are  candidates for triple systems. 
In the same way that early exoplanet discoveries taught us the 
importance of migration, analysis of multiple systems  points to 
{\it dynamical evolution\/}  in which a third star plays a special role. 
 One example of the consequences of 
a triple system is that  spectroscopic
binaries with $P< 3$~d are much more likely to have a third component
than those with $P> 12$~d (Tokovinin et al.\ 2006).  
This implies that dynamical evolution has shortened
the periods of the first group.    Conservation of 
total system angular momentum requires that the wider tertiary components move 
to wider separations. 
Scenarios involving a triple system 
have also been proposed to explain close eccentric  binaries (Perets \& Kratter
2012)  and close binaries with small mass ratios  (Moe \& Di Stefano 2015),
both required for exotic end-stage objects. 

The present study is Paper IV in  a series aimed at determining
the binary/multiple 
properties of Cepheids.  They are reasonably
massive stars (typically 4 to $8\,M_\odot$ stars, abbreviated to $6\, M\odot$
below) in a well-known evolutionary state 
with well-studied distances and reddenings.  
This study builds on 
a {\it Hubble Space Telescope\/} ({\it HST}) 
Wide Field Camera 3 (WFC3) imaging survey of 
70 Cepheids.  Evans  et al. (2013; Paper I)
 demonstrated the technique using Cepheids with reasonably massive 
companions (q = M$_2$/M$_1$ $\geq$ 0.4).  
In Papers~II and III (Evans et al.\ 2016a and 2016b), we used the WFC3 
images in the F621M and F845M filters to 
  identify  candidate resolved companions as close as approximately 
0$\farcs$5.    With this we can identify main-sequence companions typically 
as close as 300 AU.  Paper II discusses possible companions $\geq$ 5$"$ from
the Cepheid; Paper III discusses closer companions identified after
point spread function (psf) correction.  

Possible resolved companions were identified in Paper II  as follows.  A 
color-magnitude (CMD) was formed for stars  from 
the  F621M and  (F621M--F845M) data, which transform well to V and 
(V-I).  This was compared with an isochrone at the distance and with 
the reddening of the Cepheid (see Paper II).  Stars in the field
within 2 $\sigma$ of the isochrone with colors hotter than M stars 
were considered candidate resolved companions, and are listed in Paper II.    

Identification of resolved companions, particularly low-mass stars, is plagued
by contamination from galactic field stars.  However, resolved 
companions which are genuine physical companions of Cepheids (typically 50 Myr
old) will  be young.  Stars this young are active X-ray producers as 
compared with old field stars.  In the present study, we have observed a number
of possible resolved companions from the   {\it HST\/} survey with the  {\it
XMM-Newton\/} X-ray  satellite ({\it XMM\/}) to identify X-ray
active  young stars, likely to be genuine bound companions.  

Thus, the aim of this study is to identify the widest companions of Cepheids
down to very  small masses.  One goal is to determine 
the maximum separation which 
set by star formation plus subsequent dynamical
evolution.  A second aim is related to the fact that a wide system  may
easily  contain a closer binary, at least in the main-sequence phase.  A wide 
companion certainly does not guarantee a three-star system, but it is a  flag that
this may be the case.  As discussed above this opens a
number of possibilities  for its evolution.



The subsequent sections in this paper  discuss the observations and data
analysis (\S\ref{obs.dat}),  the resolved companions, upper limits  and
detections (\S\ref{res.comp}),  results (\S\ref{results}), discussion 
(\S\ref{discuss}), and a summary.

\section{Observations and Data Analysis}\label{obs.dat}

Our {\it XMM\/} target list was selected from a compilation of possible resolved
companions found in  our {\it HST\/} optical imaging survey.  Discussion of the
results of the survey is done in two  parts, determined by data-reduction
challenges (Papers II and III).  
Possible companions at $\geq\!
5\arcsec$ are good targets for {\it XMM\/} imaging, since this is  approximately
the size of the {\it XMM\/} psf. Specifically, our target list was drawn from
Table A1 in Paper II of such candidates, which would be resolvable by {\it
XMM}.   Table~\ref{xmm.obs} lists the observed Cepheids, the date of 
observation and UT at the beginning of the integration, the exposure duration
for the MOS1 CCD camera,
the reddening and distance (from Paper II),  and the {\it XMM\/}  filter used.
In addition to {\it XMM\/} observations made for this study, two  observations
were obtained from the {\it XMM\/} archive: U Sgr (PI: Motch) and $\ell$~Car
(PI: Guinan).  Of the 39 Cepheids with candidate  resolved  companions
$\geq$5$\arcsec$  (from the 70 WFC3 observations), altogether 14 of them (36\%) 
were observed with {\it XMM}.  


Data analysis was carried out as in Pillitteri et al.\ (2013), using standard
{\it XMM\/} data reduction 
(SAS)  tasks to filter events so that events in the band 0.3 to 8.0 keV were
used. Only good time intervals were included  and  high-background intervals
were removed. Source detection and upper-limit calculation were done using a
wavelet deconvolution  algorithm, as implemented in the code originally written
for {\it ROSAT\/} images  (Damiani et al.\ 1997a, 1997b) and
adapted for {\it XMM\/} images. 




\section{Candidate Resolved Companions}\label{res.comp}

The results of our {\it XMM\/} observations of possible resolved companions are
summarized in  Table~\ref{xmm.comp}.   The  successive columns are:  the
companion identification number for each Cepheid, $V$ and $V-I$ from Paper II,
the separation in arcsec, the position angle,  the separation converted to AU
using the distance from Table~\ref{xmm.obs}, the J2000 RA and Dec of the
companion, and whether an X-ray source was detected at that  location
(Y/N/?). The coordinates were measured from the WFC3 images.

\begin{figure*}
\includegraphics[width=\columnwidth]{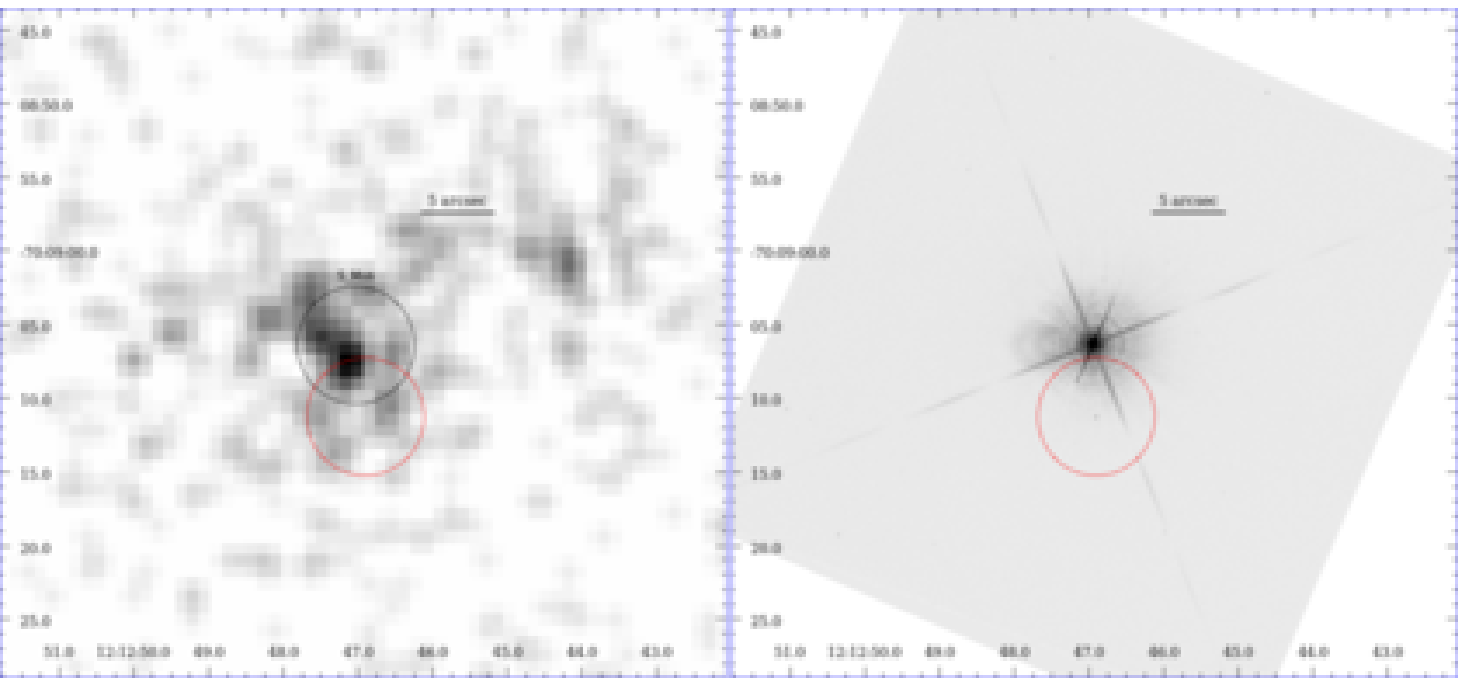}
\caption{a. (left) The {\it XMM\/} image of S Mus. The circles show the 
location of the Cepheid (black) and the possible companion (red). 
The orientation of both figures is the same with N up and E on 
the left.  The scale is indicated by the 5$\arcsec$ line.  A 
log stretch is used to emphasize faint features.  Circle sizes in both a 
and b are arbitrary. b. (right) The {\it HST} WFC3 I image of S Mus.  
The possible companion is 
circled. 
 \label{smus}}
\end{figure*}

In only three cases (S Mus, R Cru, and S Nor \#4) is there possible X-ray flux
at the location of the  resolved companion,  as shown in
Fig~\ref{smus}, Fig~\ref{rcru}, and Fig~\ref{snor}.     In two further cases
(V659 Cen and V473 Lyr) a source was detected at a position indistinguishable 
from the Cepheid itself.  These will be discussed in \S3.2.

\subsection{Detection Upper Limits}

The {\it XMM\/} exposure times were set with the aim of detecting main-sequence
companions as cool as spectral type K, which would have  $\log L_X$ of 29.2 erg
sec$^{-1}$ or greater (at the distance and with the reddening of each Cepheid).  X-ray
fluxes for M dwarfs become much fainter as temperature decreases, and exposure
times become much longer.  X-ray flux, of course, depends on the  age of the
stars.   The $\alpha$ Per cluster is approximately 50 Myr old,  making it an
appropriate comparison for Cepheids which have about the same age.  {\it
ROSAT\/} observations of the cluster were discussed by Randich et al.\ (1996). 
At approximately  this limit, they detected 88\% of K stars in their list.
Pillitteri et al.\ (2003)   use  {\it ROSAT\/} data to obtain the X-ray
luminosity distributions of G, K, and M stars for  the $\alpha$ Per cluster. 
 The X-ray detection rate at $\log L_X$ of 29.2 erg sec$^{-1}$ is 80-90\% 
for G and K stars, but falls off for M dwarfs.  A deeper {\it XMM\/} observation
of the $\alpha$ Per cluster  was discussed by Pillitteri et al.\ (2013).  Their
luminosity function found a similar detection rate. (Note that the
relevant bin in there discussion 
includes early M stars as well as G and K stars).   Because of the dependence of
X-ray flux   on both age and spectral type, and, of course, the X-ray
variability cycles of late-type stars, there is some imprecision in the
estimate.  However, this luminosity limit should ensure that most
late-type stars through the K range (to 0.5  $M_\odot$)  will be detected.  

To quantify the non-detections in Table~\ref{xmm.comp}, we have computed upper
limits  at the positions of each of the candidate companions.   Since we know
the location of the possible companion stars, we used a  3$\sigma$ detection
limit for the upper limit.   We ran the wavelet algorithm with this threshold 
to determine the level of count rates for a point-like source at the  positions
of the Cepheid companions.
Table~\ref{xmm.comp.ul} (Col.~3)  lists the upper limit.  The detections and
upper limits were obtained using both  MOS and pn images, using the standard 
relative efficiency MOS/pn cameras of 800/260 from the effective areas in the
appropriate energy range. For V473 Lyr  the count-rate limit is measured
slightly further from the Cepheid than  the position shown in
Fig.~\ref{v473lyr},  to decrease flux  from the position of the Cepheid
(although there is still likely to be some contribution from the source  at the
Cepheid position).  The  ratio of $N_H$ to $E(B-V)$ used to derive the column
densities in Col.~4 was   $5.9 \times 10^{21}\rm cm^{2}  mag^{-1}$ .  The
corresponding flux (Col.~5) was generated for this $N_H$ using an APEC plasma
emission code in the PIMMS flux calculation software 
\footnote{\tt http://cxc.harvard.edu/toolkit/pimms.jsp}
 with $\log T = 7.0$ and solar abundance for XMM using the filter in
Table~\ref{xmm.obs}.  The upper-limit X-ray luminosity (Col.~6) was then derived
using the distance.  



In summary,   most of the upper limits (Table~\ref{xmm.comp.ul}) 
are close to the
goal of  $\log L_X$ of 29.2 erg s$^{-1}$, showing that we should have  
detected most of the late F, G, and K
stars which are young and at the distance of the Cepheids. 







\subsection{X-ray Detections}

The candidate optical companions targeted by our {\it XMM\/} observations are
$\geq$5$\arcsec$ from the Cepheid, allowing us to avoid confusion with X-ray
emission from the Cepheid itself.  In five cases, we detected an X-ray source 
 of which four are at
or near the Cepheid itself. These detections
are listed in   Table~\ref{xmm.det}, which gives
fluxes, luminosities, and comments about  the sources.  The detections of  each
source are discussed below.  For comparison, several Cepheids (Polaris, $\delta$
Cep, and $\beta$ Dor) have been detected in X~rays with typical  luminosities
of   $\log L_X = 28.6$--29.0 erg s$^{-1}$. The {\it XMM\/} observation of  
$\beta$ Dor has a  $\log L_X$ of 29.0 ergs s$^{-1}$, and a soft spectrum (Engle
et al.\ 2009; Engle et al. 2014). 
Polaris was observed with {\it Chandra\/} (Evans et al.\ 2010)
and has  $\log L_X$ = 28.9 erg s$^{-1}$, and also a soft spectrum.  As they
discussed, the interpretation is complicated by the fact that Polaris has a
low-mass companion which could also produce or contribute to the X-ray flux.  
$\delta$ Cep itself has {\it XMM\/} observations at several phases.  The flux 
is reasonably constant  at a luminosity of about $\log L_X$ = 28.6 erg
s$^{-1}$.   One point is significantly brighter, suggesting variability with
phase, which would  indicate that the flux is produced by the Cepheid.  However,
recently high-accuracy  radial velocities show low-level orbital motion,
consistent with a low-mass  companion (Anderson et al.\ 2015). As with Polaris,
the companion could contribute some or all the X-ray flux.  The luminosities of
these three systems provide either detections or upper limits to the X-ray
luminosity of the Cepheid itself, which is below the luminosities in
Table~\ref{xmm.det}

In this section we discuss each of the {\it XMM\/}
 detections in Table~\ref{xmm.det}.  


{\bf S Mus}:  A possible companion of S Mus lies within the psf of the  {\it
XMM\/}  image (Fig~\ref{smus}).  From the image we cannot tell whether the
X-rays  come from the companion  or from the Cepheid itself. We have recently
obtained an observation of S Mus with the {\it Chandra} X-ray satellite.  From 
the image showing that the X-rays are not produced by 
the resolved companion, but from the location of the Cepheid.  Full discussion
of the results is in preparation.  S Mus is a
well-known binary system with a period of 1.38 years (Evans et al.\ 2006, and
references  therein).  Velocities of the hot companion have been measured in the
ultraviolet (B\"ohm-Vitense et al.\ 1997) and show no sign that the companion is
itself a binary.  The companion has a spectral type of B3 V, and is 
the one companion among the XMM observations
hot enough that it could produce X-rays itself, which seems the most likely
source of the X-rays.

\begin{figure*}
\resizebox{\columnwidth}{!}{
\includegraphics{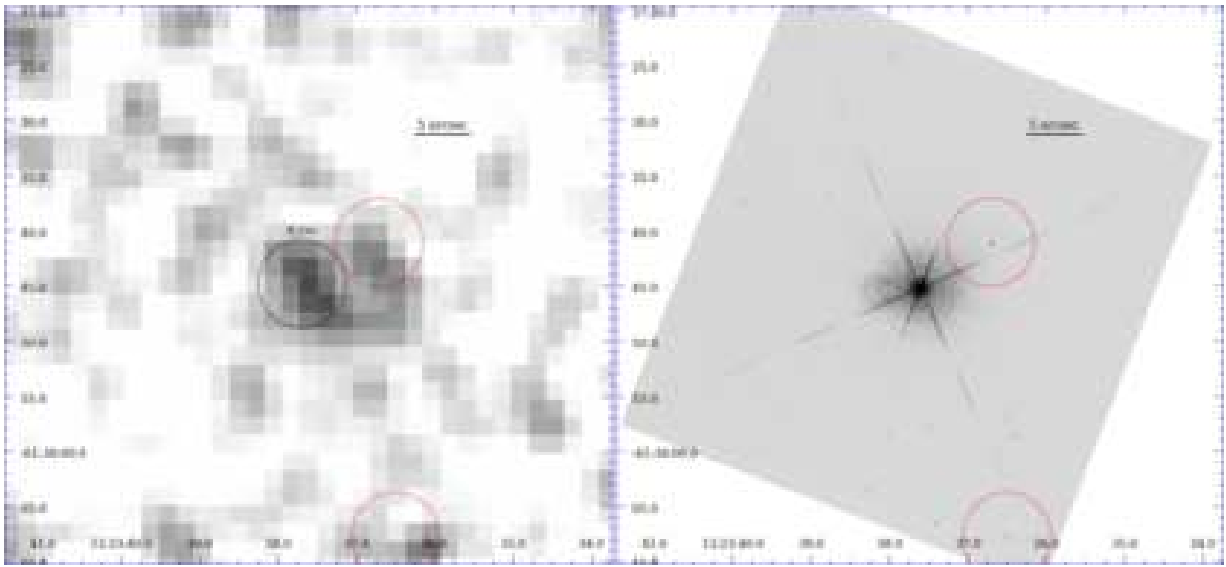}
}
\caption{a. (left). The {\it XMM\/} image of R Cru with circles to show the locations 
of the Cepheid and two possible companions.  b. (right) The {\it HST} WFC3 I 
image of R Cru.  Both possible companions
are circled.  The image orientation and treatment are the same as for 
Figure~\ref{smus}.
 \label{rcru}}
\end{figure*}
\begin{figure*}
 \includegraphics[width=\columnwidth]{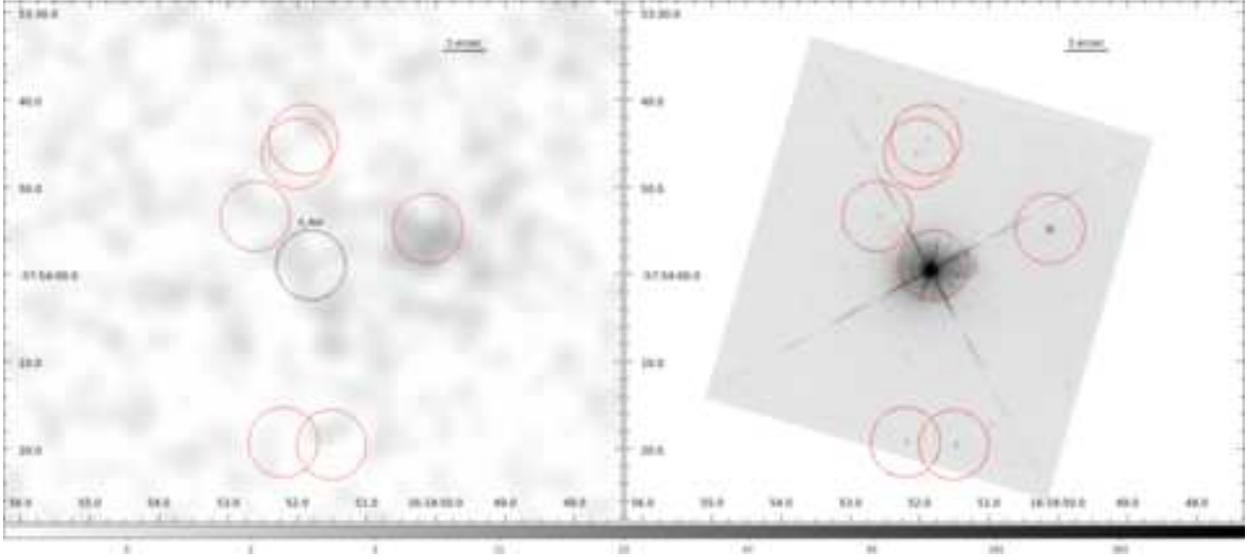}
\caption{a. (left). The {\it XMM\/} image of S Nor with circles to show the locations 
of the Cepheid and the possible companions.  b. (right) The {\it HST} WFC3 I 
image of S Nor.  Only one possible companion was detected in XMM image, 
the one the furthest to the W. 
  The image orientation and treatment are the same as for 
Figure~\ref{smus}.
 \label{snor}}
\end{figure*}

{\bf R Cru}: As with S Mus, the {\it XMM\/}  image (Fig~\ref{rcru}) cannot
determine whether the X-rays come from the close (8$\arcsec$) companion or the
Cepheid.  In  addition, in Paper III, we find  a  possible companion 
1$\farcs$9 from the Cepheid. Either of  the companions or the  Cepheid could
produce the X-rays.  Fig.~\ref{rcru} also shows that there is  a small
brightening in X-rays at the position of the resolved companion to the right of
the Cepheid.  We have examined that area carefully, and there  is {\it not\/} a
significant source at that location, 
but rather a fluctuation  similar to other
background fluctuations apparent in the figure.  

\begin{figure*}
\includegraphics[angle=270,width=\textwidth]{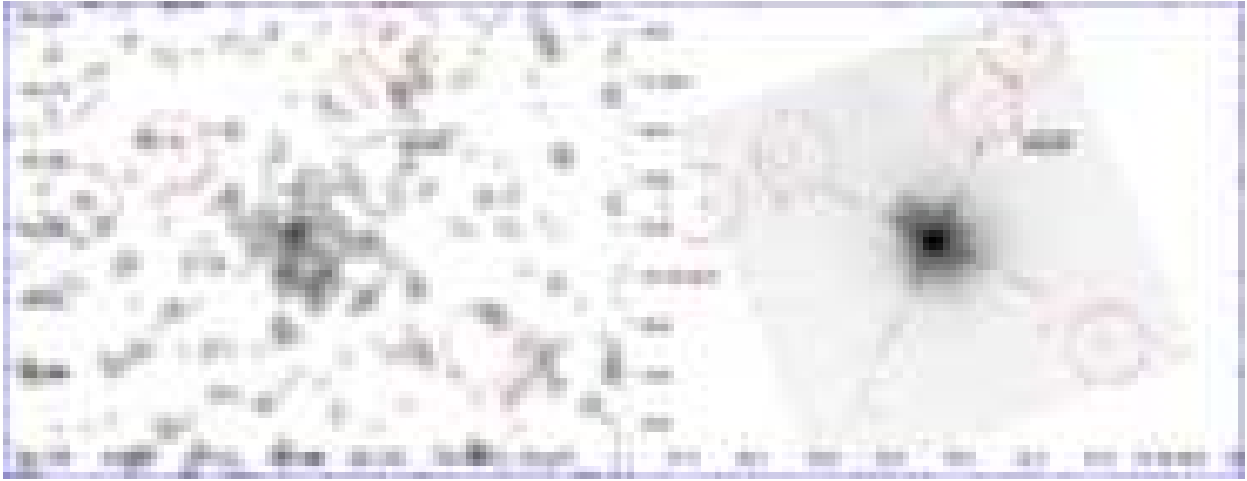}
\caption{a. (left). The XMM image of V659 Cen with circles to show the locations 
of the Cepheid and five possible companions.  b. (right) The HST WFC3 I 
image of V659 Cen.  All possible companions
are circled.  The image orientation and treatment are the same as for 
Figure~\ref{smus}.
 \label{v659cen}}
\end{figure*}

\begin{figure*}
 \includegraphics[width=\textwidth]{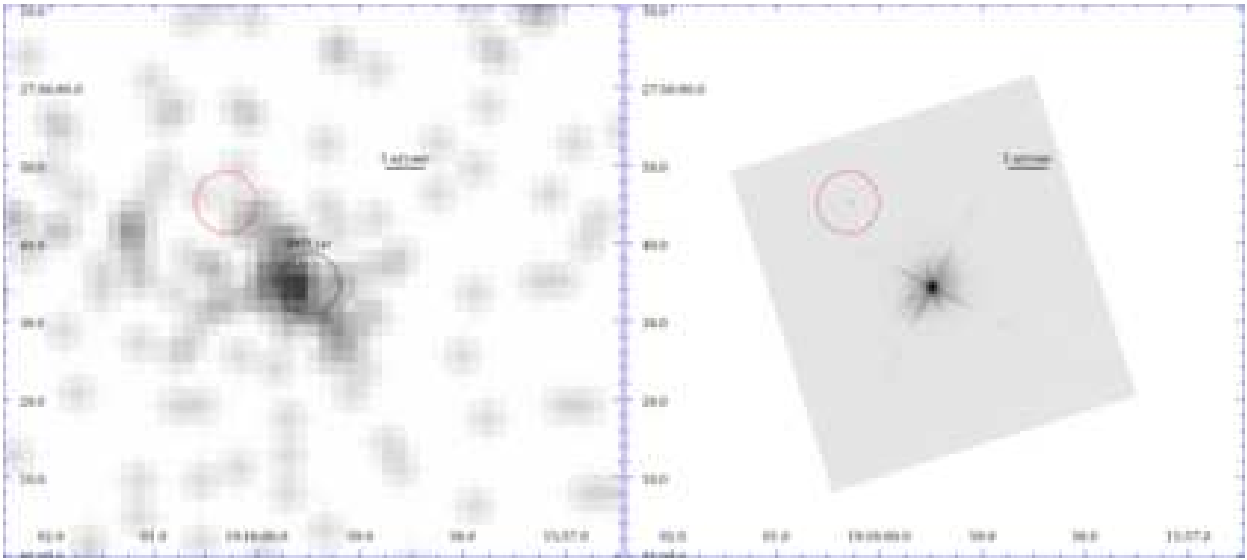}
\caption{a. (left). The XMM image of V473 Lyr with circles to show the locations 
of the Cepheid and the possible companion.  b. (right) The HST WFC3 I 
image of V473 Lyr.  The possible companion
is circled.  The image orientation and treatment are the same as for 
Figure~\ref{smus}.
 \label{v473lyr}}
\end{figure*}

{\bf V659 Cen}:  None of the five resolved possible companions in
Fig~\ref{v659cen} are  coincident with an X-ray source.  However, in Paper III,
there is  a  possible companion   at about 0$\farcs$7 from the Cepheid.  Either
it or the  Cepheid could produce the detection. 

{\bf V473 Lyr}:  The X-rays in  Fig~\ref{v473lyr} come from the vicinity of the
Cepheid rather than the possible companion (separated by 15\arcsec).  This
raises the question  whether there is a closer low-mass  companion which could
produce the X-rays. There are considerable data which address this topic.   In
Paper III there is no sign of a close companion.  Conclusions from velocities
are tricky because of the variable pulsation amplitude. The star has been 
discussed recently by Molnar \& Szabados (2014), who conclude that it does not
show binary motion, based largely on the 5 years of velocity data from Burki
(1984). A binary companion can never be completely ruled out, but there is no
evidence for  one from either velocities or imaging. Thus, the Cepheid itself is
a possible source. This star is one of the most unusual Cepheids known, unique
among classical  Cepheids in having a large-scale amplitude variation,
resembling the Blazhko  variations in RR Lyrae stars.    

{\bf S Nor \#4}:  This is the only X-ray source that is unequivocally from the 
resolved companion (Fig~\ref{snor}).  However, since the Cepheid is a member of a 
populous open cluster, there is a significant chance of an alignment with a 
cluster star. 


In Table~\ref{xmm.det} we include the median energy for each of the detections. 
The  sources on the images are relatively weak, so full spectral analysis is not
feasible. However, the median energy provides information which can help
distinguish between  a low-mass star and a cool supergiant, which has a softer
spectrum.  As an example, see Fig.~3 in Evans et al.\ (2010), which compares 
the Polaris spectrum with the
stacked spectrum from Orion Nebula low-mass stars (Feigelson et al.\ 2005).  The
low-mass stars have a sharp energy peak  at 1 keV, with a tail extending out to
2 keV.  Stars with earlier spectral type (as  well as older stars), including
the Polaris system, have energy peaks at about 0.9 keV.  


The combination of the  median energy values and $L_X$ in  Table~\ref{xmm.det} 
provides the following information about the X-ray sources.  S Nor \#4 is the
only undisputed X-ray source resolved  from the Cepheid.  Its median energy
(0.95 keV) and $L_X$ are consistent with a low-mass star.   In fact, the  $L_X$
for all the detected sources is consistent with low-mass stars, but 
S Mus is the outlyer in the high end.   For the four
systems where the position of the X-ray source is indistinguishable from the Cepheid,  R
Cru is the star with the highest median energy (1.24 keV).  This is a strong 
indication that the X-rays are produced by the close (1$\farcs$9) late-type
companion with a harder spectrum than a supergiant.   On the other hand, the
much softer median energy for V473 Lyr points to either the Cepheid itself or an
as yet unknown companion earlier than G spectral type.   For 
V659 Cen a low-mass companion is consistent with both the median
energy and $L_X$, and is there other  evidence of such a
companion.   For S Mus, the median energy and L$_X$ are consistent with X-rays 
produced by shocks in an outflowing wind, as might be produced in the hot 
companion (see, for example, Stelzer, et al. 2005)

\begin{figure*}
 \includegraphics[angle=270,width=\textwidth]{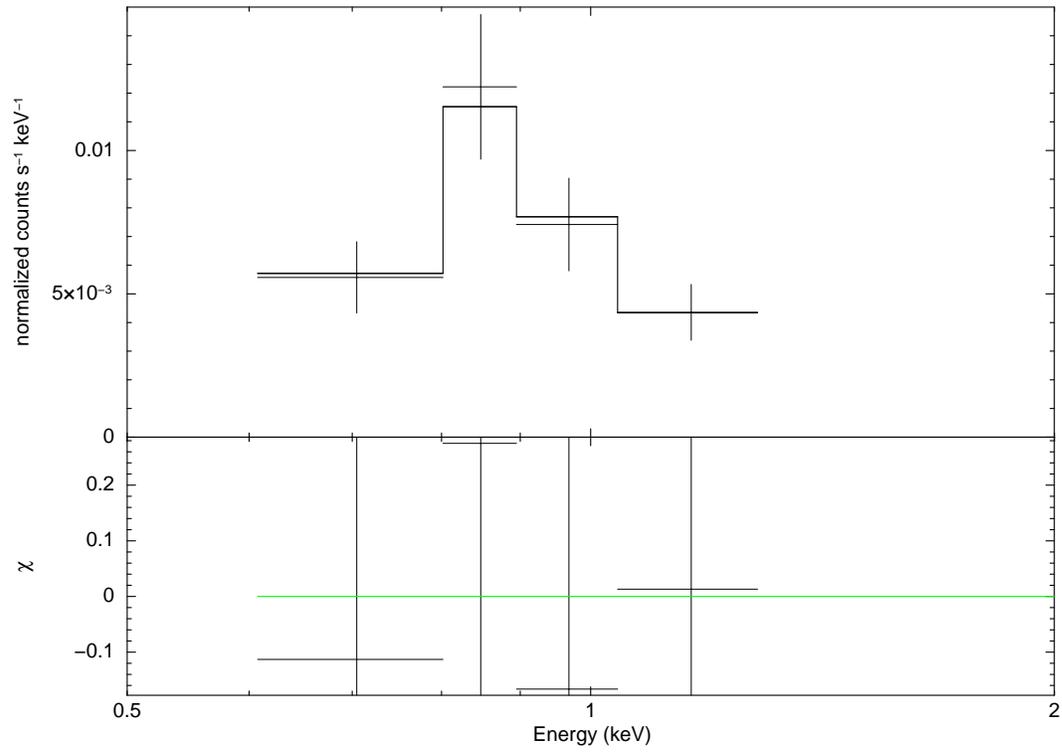}
\caption{The spectrum of the S Mus source.  Top: normalized counts: +; 
histogram spectral fit.  Bottom: Differences between data and fit (normalized 
by the s.d. of the bin).
 \label{smus.pnspec}}
\end{figure*}

Only the S Mus source has enough counts to make it reasonable to look at the 
spectrum (Fig.~\ref{smus.pnspec}).  Although the counts  warrant only crude 
energy bins, energy maximum  is clear.   The resulting energy from the fit ($kT
= 0.64 \pm 0.14$ keV) is consistent with  a low-mass star, although the $N_H$  
($(5 \pm 2) \times 10^{21}$ cm$^{-2}$) and hence the unabsorbed flux ($6.3\times
10^{-14}$ erg cm$^{-2}$ s$^{-1}$ in the 0.3-8.0 keV band) is  larger than that
in Table~\ref{xmm.det}.

In summary, of the five {\it XMM\/} detections, two stars (R Cru and V659 Cen)
have a close  companion ($<$2$\arcsec$ from the Cepheid) which 
are the most likely sources of  the
X-rays.  For S Mus), the working model is that the X-rays are probably produced 
by the B3 companion.  For S
Nor, the late-type resolved companion is detected, though  whether it is
gravitationally bound to the Cepheid or a cluster member is unknown.  For the
final detection, V473 Lyr, the only candidate for the X-ray source appears to be
the Cepheid.


We stress that the main goal of this study was to search for X-rays from 
young low-mass stars to identify Cepheid companions.  These are
coronal stars, for which the X-ray properties have been well studied. 
The mechanism for producing X-rays in Cepheids 
themselves is not  well understood.
They are also coronal stars.  One possible mechanism is pulsation-driven 
shocks, as discussed by Engle, et al. (2014).  Another possibility is 
collisionless shocks (Ruby, et al. 2016).

\section{Results}\label{results}

\begin{figure*}
 \includegraphics[angle=270,width=\textwidth]{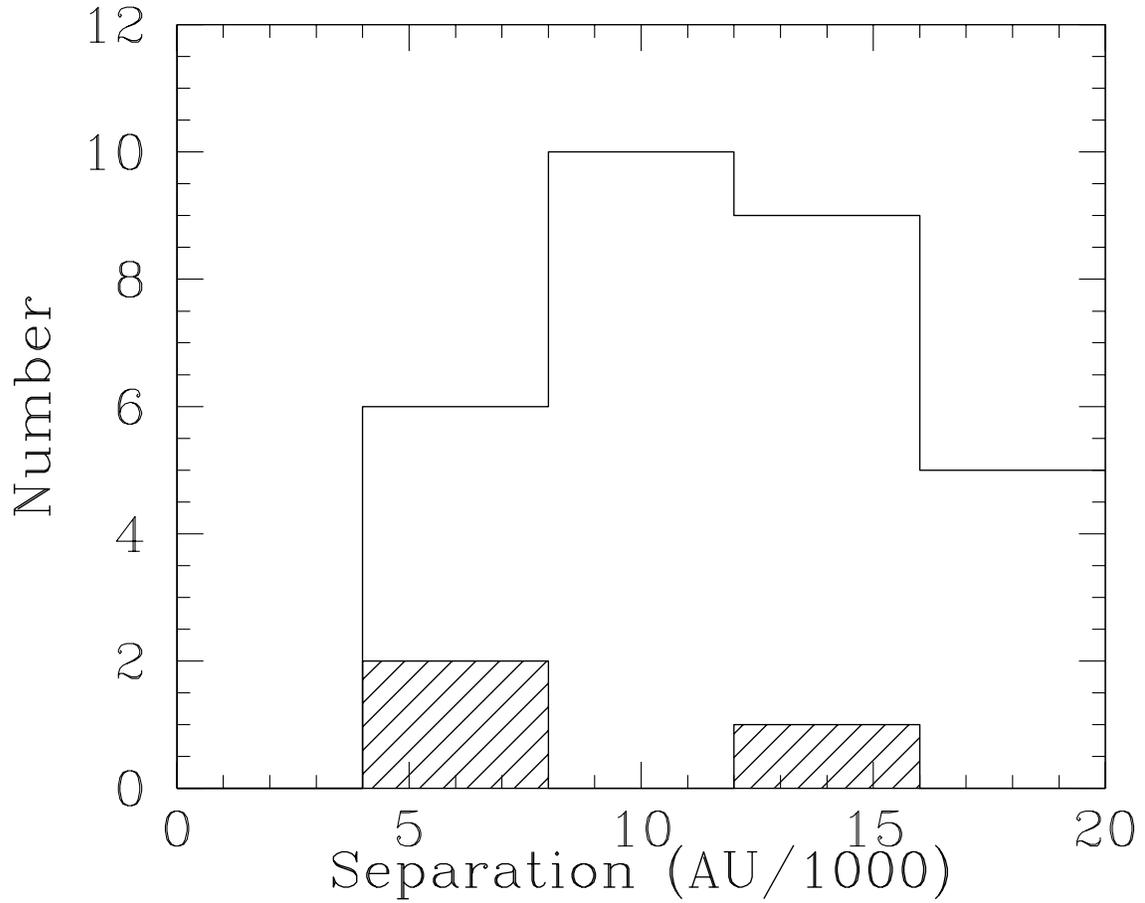}
\caption{The separations from the Cepheids of the possible companions
in  Table~\ref{xmm.comp}.  Those with possible X-ray detections are 
hatched. Among the detected sources,  S Nor \#4 is the star with a 
separation of 13300 AU, thought
to be a cluster member.  The two detected sources with the smallest separations
(S Mus 3950 AU and R Cru 6330 AU) are both thought to come from closer 
companions than those in this figure.  See text for discussion.
 \label{sep}}
\end{figure*}

The distribution of separations for the possible companions investigated in 
Table~\ref{xmm.comp} is shown in Fig~\ref{sep}.   Appropriate to  {\it XMM}'s
spatial resolution, the Cepheids observed all have possible companions
at separations $\geq$5$\arcsec$.  
 In   Fig~\ref{sep}, the optically resolved
candidate companions are  overwhelmingly {\bf not\/} found  to have the X-ray
strength expected for a low-mass star about 50 Myr old.  The star which is the
exception to the finding that  there are {\bf no} companions in the survey at greater
distances than 8000 AU is the companion to S Nor.  This is one of two stars in
the survey (S Nor and U Sgr) which is a member of a  populous open cluster. 
This increases the likelihood of a chance  alignment with a cluster member.  We
thus consider the S Nor \#4 companion
at a projected  separation of 13,300 AU, to be an
outlier,  and most likely not to be   gravitationally bound to the Cepheid.

The {\it HST} WFC3 survey covered 70 Cepheids, of which 39 were found to have 
possible resolved companions with separations $\geq$5$\arcsec$. We observed 
14 (36\%) of the possible companions
 with {\it XMM} and found no young stars (with the exception of S Nor \#4, 
which is most probably a cluster member).  
Thus, we have found no probable resolved companions with  
separations larger than 3950 AU (S Mus, Table 2).   

\section{Discussion}\label{discuss}

Systems containing Cepheids have undergone several periods of reorganization.  
As massive stars they are highly likely to be found in binary or multiple 
systems.  Much of the dynamical adjustment from the initial configuration  in a
system of three or more members happens quickly and many will have already
produced a hierarchical system by the time the primary reaches the zero-age
main  sequence (ZAMS).  As stressed above, however, triple (and higher) systems can 
still evolve dynamically.     This can result in the ejection of a star
(typically the least massive) from the system.  

After the ZAMS, since Cepheids are post-red-giant stars, very
close binaries will have 
undergone Roche-lobe overflow.  In extreme cases this results in mergers.
Cepheids which began life as B stars have in some cases had their 
binary properties altered in this way, since there are no Cepheid binaries 
with shorter periods than a year in the Milky Way (Sugars \& Evans 1996) 

We stress three features of the current investigation which contribute to  the
understanding of binary/multiple properties in fairly massive stars (typically
$6\, M_\odot$).  First, we have typically  searched
an area as wide as the canonical  0.1 pc (20,000 AU) 
 in our {\it HST\/} images for  possible companions, 
which is the extent expected for which physical companions.    Second, the
{\it XMM\/} observations provide a severe  winnowing of the candidate list, and
an important  constraint on the extent of  bound companions.  Third, the
magnitude difference between the Cepheid and a possible companion can be more
than $\Delta V = 10$ mag, including companions through K stars. A late K star
would have a mass ratio of 0.1 with a $6\, M_\odot$ Cepheid.    


Full discussion of the results of  closer companion will be provided in Paper III; 
however we will make a preliminary
comparison with other recent results, particularly of the extent of wide  
systems. The  recent survey of  resolved  O star companions  (Sana et al.\ 2014;
SMASH) plans a further analysis of binary properties; however, a preliminary 
comparison with their results   is warranted at this stage.  The VLT and VLTI
instruments they used cover from  0.5  through 10$^{4}$ mas, (1 AU to 0.1 pc at
the typical distance of their O stars
of 2 kpc).  (The typical distance in our Cepheid survey is
700 pc).  It is only their widest separations which are  comparable to the
Cepheids discussed here.   An interesting change occurs for their detections  at
about 1$\arcsec$.  The magnitude differences are larger (up to 8 mag in $H$);
however while  there are many comparatively faint companions, the number of
bright companions  drops off as compared with closer regions.  Indeed, their
probability analysis also  concluded that the low-mass companions at separations
$\geq$2$\arcsec$ are not  physical companions.   For a $20\, M_\odot$ system,
the separation of $\sim$2000 AU where this happens is  similar to the widest
separation we find.

Tokovinin et al.\ (2006) made a survey of 165 solar-type spectroscopic  binaries
to search for wider components.  They noted a decrease in third  components at
periods $>$10$^5$ years ($\simeq$ 3000 AU).   In a more recent distance-limited 
survey of solar-mass stars, Tokovinin (2014) finds a decrease in the  widest
companions at about the same period.  A similar result was found by  Raghavan et
al.\ (2010), also for solar-mass stars.



\section{Summary}

We discuss {\it XMM\/} observations of possible resolved companions of Cepheids
from an {\it HST} WFC3 survey.  
X-ray observations differentiate between young, low-mass stars (physical companions 
of Cepheids) and old field stars.  

$\bullet$  Of
the possible resolved companions, only one was unambiguously 
detected, implying that it is a 
young star.   The Cepheid involved is S Nor \#4, which is a member of a well-populated
cluster.  For this reason, and because  the companion   is separated from the
Cepheid by as much as  15$\arcsec$ (13,300 AU), it is likely that this is a 
cluster member rather than a gravitationally bound companion.  

$\bullet$ Two of the
Cepheids  (S Mus and R Cru) have an X-ray source which is not resolved from 
the Cepheid on the {\it XMM} images.  However, on a {\it Chandra} image of S Mus, 
the X-rays are shown to be produced by the Cepheid/spectroscopic 
binary.    

$\bullet$ R Cru and also V659 Cen have
additional  companions closer than the  limit of 5$\arcsec$ for this study,
which are the likely X-ray sources (to be discussed in Paper III).   

$\bullet$ One final X-ray detection (V473 Lyr) has
no known companion; hence the prime  suspect is the Cepheid itself.  It is a
unique Cepheid with a variable amplitude.  

$\bullet$ The 14 stars observed with {\it XMM\/} constitute 36\% of the candidate
companions.  None were found to be young likely physical companions with 
a separation $\geq$5$\arcsec$ or wider than a projected separation of 4000 AU
(assuming that S Nor \#4 is a cluster member).

The full binary frequency will be discussed in  Paper III 
including the entire {\it HST\/} WFC3 survey.

\acknowledgments

 Support for this work was also provided  from the {\it Chandra} X-ray Center NASA 
Contract NAS8-03060.  Funding was provided from HST GO-13369.001-A (to NRE and
IP).  Vizier and  SIMBAD were used in the preparation of this study.
Comments from an anonymous referee improved the clarity of the presentation.

\begin{deluxetable}{lccccc}
\tablecaption{{\it XMM\/} Observations }
\tablewidth{0pt}
\tablehead{
\colhead{Star} & \colhead{Date \& UT} &  \colhead{Exp.}&  \colhead{$E(B-V)$} &  \colhead{Distance} & \colhead{Filter} \\ 
\colhead{} &   \colhead{ } & \colhead{MOS1 [ks]}&  \colhead{} &  \colhead{[pc]} & \colhead{} 
}
\startdata
$\ell$ Car & 2010-02-08 10:32:32 & 52.9  & 0.17  &   506 & Thick   \\ 
V659 Cen   & 2013-09-07 20:10:37 & 20.8  & 0.21  &   753 & Medium  \\  
V737 Cen   & 2014-01-26 14:29:20 & 31.7  & 0.22  &   848 & Medium  \\ 
R Cru      & 2014-01-04 19:56:55 & 22.0  & 0.19  &   829 & Medium  \\ 
S Cru      & 2013-08-20 21:52:30 & 11.4  & 0.16  &   724 & Medium  \\ 
X Cyg      & 2013-04-26 02:46:47 & 30.9  & 0.29  &   981 & Medium  \\ 
V473 Lyr   & 2013-09-22 09:49:34 & 6.6   & 0.03  &   553 & Medium  \\ 
R Mus      & 2013-02-15 01:19:42 & 17.4  & 0.12  &   844 & Medium  \\ 
S Mus      & 2013-01-05 14:36:48 & 25.3  & 0.21  &   789 & Medium  \\
S Nor      & 2015-03-13 09:10:05 & 33.1   &  0.19  &  910 & Medium \\
Y Oph      & 2012-09-12 06:52:49 & 9.5  &  0.65  &  510 & Medium \\ 
V440 Per   & 2013-09-02 21:48:18 & 24.4  &  0.27  &  791 & Medium \\ 
U Sgr      & 2006-10-11 23:29:34 & 29.5  &  0.40  &  617 & Medium \\ 
Y Sgr      & 2013-09-29 04:56:27 & 13.2  &  0.20  &  505 & Medium \\ 
\enddata
\label{xmm.obs}
\end{deluxetable}



\begin{deluxetable}{ccccccccc}
\tablecaption{Candidate Resolved Companions}
\tablewidth{0pt}
\tablehead{
\colhead{Companion} &
\colhead{$V$} & \colhead{$V-I$} &  \colhead{Sep.} &  \colhead{PA} &  \colhead{Sep.} &  \colhead{R.A.} &  \colhead{Dec.}   
&  \colhead{X-ray}  \\ 
\colhead{No.} &
\colhead{} &   \colhead{} & \colhead{[arcsec]} & \colhead{[$^\circ$]} & \colhead{[AU]} & \colhead{[J2000]} & \colhead{[J2000]}
 & \colhead{Source?}  
}
\startdata
\multicolumn{9}{c}{$\ell$ Car} \\
1   &  15.21   &     1.09   &     19.1   &      9.7  & 9,660 & 09 45 15.3 & $-$62 30 09.5     & N  \\
\multicolumn{9}{c}{X Cyg} \\
1   &  18.61   &     1.68   &     12.9   &  96.2 & 12,700  &  20 43 24.4 & +35 35 28.8     & N  \\
2  & 16.33   &     1.36   &     14.8   &     298.3 & 14,500  &  20 43 23.5 & +35 35 03.9     & N \\
\multicolumn{9}{c}{V659 Cen} \\
1   &  15.87   &     1.20   &     22.1   & 335.3     & 16,600 & 13 31 32.1 & $-$61 34 36.3     & N     \\
2  &  18.09   &     1.59   &     14.7   &  340.7 & 11,100  &   13 31 32.7 &  $-$61 34 42.6     &   N  \\
3  &  16.57   &     1.27   &     20.2   & 238.5 & 15,200 & 13 31 30.9 & $-$61 35 07.1     &   N    \\
4  &  15.73   &     1.17   &     23.8   & 81.1 & 17,900 &  13 31 36.7 & $-$61 34 52.8     &  N  \\
5  &  17.59   &     1.47   &     17.0   &  58.6 & 12,800 &  13 31 35.4 & $-$61 34 47.6      & N  \\
\multicolumn{9}{c}{V737 Cen} \\
1   &  17.22   &     1.61   &      7.3   &  294.9 & 6,190 & 14 37 11.0 &  $-$62 00 36.1     &  N    \\
2   &  17.67   &     1.61   &     17.1   &  231.1  & 14,500 &  14 37 10.0 & $-$62 00 49.9     & N  \\
\multicolumn{9}{c}{R Cru} \\
1  &  16.28  &     1.17   &	 7.64   & 301.7   & 6,330 &  12 23 36.7 & $-$61 37 41.1     &  ? \\
2 &  17.94   &	1.45     &    23.92      & 199.5     &  19,800        &   12 23 36.5 & -61 38 07.8     &  N \\
\multicolumn{9}{c}{S Cru} \\
1  & 16.59   &     1.58   &     13.8   &  70.4 & 9,990 &  12 54 23.6 & $-$58 25 45.5     &  N    \\
2  &  17.90   &     1.55   &     11.9   &  20.0 & 8,620 &   12 54 22.5 &  $-$58 25 39.0     &  N   \\
\multicolumn{9}{c}{V473 Lyr} \\
1   &  14.89   &     1.23   &     15.0   & 44.1  & 8,300 &  19 16 00.3 & +27 55 45.2     &  N  \\
\multicolumn{9}{c}{R Mus} \\
1  &  15.68   &     1.17   &      6.9   & 328.1  & 5,820 &  12 42 04.3 & $-$69 24 21.5     &  N    \\
\multicolumn{9}{c}{S Mus} \\
1  &  17.94   &     1.56   &      5.0   & 182.5  & 3,950 &  12 12 46.9 & $-$70 09 11.3     &   ? \\
\multicolumn{9}{c}{S Nor} \\
1 &  16.45   &     1.15   &     19.8   &  172.0 & 18,000 & 16 18 52.2 & $-$57 54 19.3     &  N  \\
2 & 16.32   &     1.20   &     20.1   &   188.5 & 18,300  & 16 18 51.5 & $-$57 54 19.4      &  N \\
3 &  18.06   &     1.44   &      8.5   &   44.1 & 7,740  & 16 18 52.6 & $-$57 53 53.5      &  N \\
4 &   13.95   &     0.90   &     14.6  &  288.9   & 13,300 & 16 18 50.1 & $-$57 53 54.9     &  Y \\
5 & 18.00   &     1.52   &     15.0   & 1.1   & 13,600  & 16 18 51.9 & $-$57 53 44.6     &  N \\
6 &  17.37   &     1.54   &     13.5   &  6.6 & 12,300  & 16 18 52.0 & $-$57 53 46.2     &  N \\
\multicolumn{9}{c}{Y Oph} \\
1    &  17.13   &     2.00   &     18.1   & 211.9 & 9,230  & 17 52 38.1 &  $-$06 08 52.5     &  N    \\
\multicolumn{9}{c}{V440} \\
1 &  15.72   &     1.21   &     10.9   &  305.2  & 8,620 &   02 23 50.7 & +55 21 59.7     &  N  \\
2 &  13.83   &     1.01   &     10.6   &  130.6  & 8,380 &  02 23 52.7 & +55 21 46.5     &  N   \\
\multicolumn{9}{c}{U Sgr} \\
1   &  16.31   &     1.52   &     19.4   & 18.1   & 12,000 &  18 31 53.8 & $-$19 07 11.8      & N \\
2 & 17.42   &     1.95   &     13.9   &  126.7   & 8,580 & 18 31 54.1 & $-$19 07 38.5     &  N \\
3 &  17.66   &     1.94   &     17.1   & 163.8  & 10,500 & 18 31 53.7 & $-$19 07 46.7     & N \\
\multicolumn{9}{c}{Y Sgr} \\
1   &  17.06   &     1.85   &     10.6   & 204.2  & 5,350 &  18 21 22.7 & $-$18 51 45.7     &  N    \\
\enddata
\label{xmm.comp}
\end{deluxetable}

\begin{deluxetable}{lccccc}
\tablecaption{{\it XMM\/} Upper Limits of Candidate Companions}
\tablewidth{0pt}
\tablehead{
\colhead{Cepheid}   &  \colhead{Companion}  &  \colhead{Upper Limit}
&  \colhead{$N_H$} &  \colhead{Flux} &  \colhead{$\log L_X$} \\ 
 \colhead{}   & \colhead{No.} & \colhead{[ct ks$^{-1}$]}  
&  \colhead{[10$^{21}$ cm$^{-2}$]} &  \colhead{[10$^{-15}$ ergs cm$^{-2}$ s$^{-1}$] } 
&  \colhead{[ergs s$^{-1}$]} 
}
\startdata
$\ell$ Car  & 1 &  0.197 & 1.00 & 1.72 & 28.72  \\ 
X Cyg   & 1 & 0.739  & 1.71 &  6.66 & 29.89  \\ 
 &  2 & 0.797 &  & 7.18    & 29.92 \\
V659 Cen  & 1 &  0.230    & 1.24 &  1.82 & 29.09   \\  
 &  2 & 0.238  & & 1.89 &  29.11 \\
  &  3 &  0.234    &  & 1.86   & 29.10   \\
 &  4 &  0.229   & & 1.82  & 29.09  \\
 &  5 & 0.234  & & 1.86   & 29.10  \\
V737 Cen   & 1 & 0.358  & 1.30 &  2.89 & 29.40   \\  
 &  2 & 0.369  & & 2.98  & 29.41 \\
R Cru &  2 & 0.428   & 1.12 &  3.28  & 29.43   \\
S Cru    & 1 &  0.876  & 0.94 & 6.38  &  29.60    \\ 
 &  2 & 0.867   & & 6.31 & 29.60 \\
V473 Lyr & & 2.35\/ & 0.18 &  16.1  & 29.76   \\
R Mus  & 1 &  0.514  & 0.71 & 3.49  & 29.48  \\ 
S Nor  & 1 & 0.340 & 1.12  & 2.61 & 29.41  \\
  & 2 & 0.341 &   & 2.62  & 29.42 \\  
  & 3 &  0.359   &   & 2.75  & 29.44 \\
  & 5 &  0.375  &   &  2.88  & 29.47 \\
  & 6 &   0.374  &   & 2.87 & 29.46 \\
Y Oph &  1 & 0.420     & 3.83 & 6.31  & 29.29   \\ 
V440 Per  & 1 & 0.231  & 1.49 & 1.97  & 29.17   \\ 
  & 2 &  0.223  & &   1.90 & 29.15 \\
U Sgr  & 1 &  0.238   & 2.36 & 2.53  & 29.06  \\ 
  & 2 &  0.229  & & 2.44  & 29.05 \\
  & 3 & 0.231 & & 2.46  & 29.05 \\
Y Sgr  & 1 &  0.747  & 1.18 & 5.83   & 29.25   \\ 
\enddata
\label{xmm.comp.ul}
\end{deluxetable}

\begin{deluxetable}{lcccccl}
\tablecaption{{\it XMM\/} Detections}
\tablewidth{0pt}
\tablehead{
\colhead{Star} & \colhead{Source}  
&  \colhead{$N_H$} &  \colhead{Flux} &  \colhead{$\log L_X$} &  \colhead{Med E}  &  \colhead{Comments} \\ 
 \colhead{} &  \colhead{[ct ks$^{-1}$]}  
&  \colhead{[10$^{21}$ cm$^{-2}$]} &  \colhead{[10$^{-15}$ ergs cm$^{-2}$ s$^{-1}$]  } 
&  \colhead{[ergs s$^{-1}$]}  &  \colhead{ [keV]}  &  \colhead{} 
}
\startdata
V659 Cen &   5.97 & 1.24 & 4.74   & 29.51 & 0.88   & Cep? 0$\farcs$7 comp \\  
R Cru    &   1.00     & 1.12 & 7.67   & 29.80 & 1.24  & Cep? 1$\farcs$9 comp \\ 
V473 Lyr &   3.61 & 0.18  & 20.5  & 29.88 &  0.66  & Cep? \\ 
S Mus &   4.36   & 1.24 & 34.6  & 30.46   & 0.93 & Cep?   \\ 
S Nor \#4  &  1.32  & 1.12  & 10.1  &  30.00 &  0.95  \\
\enddata
\label{xmm.det}
\end{deluxetable}

\end{document}